\documentclass[11pt, letterpaper]{article}

\usepackage{jheppub}

\usepackage{amsmath}
\usepackage{amssymb}
\usepackage{amsfonts}

\newcommand{\ud}{\textrm{d}}
\newcommand{\uD}{\textrm{D}}

\renewcommand{\bar}{\overline}

\newcommand{\Tr}{\textrm{Tr}}

\newcommand{\vev}[1]{\left<\,#1\,\right>}
\newcommand{\uvev}[1]{\left<\,#1\,\right>^\circ}

\newcommand{\rst}[1]{\raise+.6ex\hbox{#1}}

\title{Correlation functions in the holographic replica method}
\author{Yanwen Shang}
\affiliation{Perimeter Institute for Theoretical Physics, \\
31 Caroline St. N., Waterloo, ON, Canada, N2L 2Y5}

\emailAdd{yshang@perimeterinstitute.ca}

\abstract{Disorder has long been a difficult subject in condensed 
matter systems and the The replica method is a well-known tool in this field.
Implementing the replica method the AdS/CFT correspondence 
has been proposed and discussed in literatures. We point out, for any CFT that has
a holographic dual and to the leading order of the large-$N$ expansion, 
the corrections due to the presence of random disorder to 
any connected correlation functions vanish identically, 
provided that the disorder strength is normalized as discussed in
literatures and that the symmetry among replicas is unbroken. Same must
hold true to any observables that are determined by the connected 
correlation functions through a linear relation. This behavior
resembles strongly that of a free theory where disorder is coupled to
the fundamental field. We demonstrate this by both the means of 
holographic principle and field theory analysis in a toy model. 
We also propose ways of evaluating the non-zero
sub-leading effects perturbatively in terms of the disorder
strength and discuss a novel possibility of defining a new holographic
dual if we adopt a different normalization for the disorder strength.}

\begin{document}

\maketitle

\section{Introduction}
\label{sec:intro}
Random disorder is common in many-body systems. In condensed matters, 
spatial inhomogeneity, or impurities, is almost always present. In certain
circumstances, they are the key element responsible for important physical
properties. A famous example is the DC conductivity, which is infinite
theoretically in a translational symmetric system and only
becomes finite because of the presence of impurities in real material. 
Despite its importance, disorder in quantum system 
remains a difficult problem, particularly when the system is strongly coupled.

One often treats the impurities statistically. It is first introduced into
the theory as a classical source coupled to a certain operator in a QFT, and
then the system is averaged over a probability distribution functional for
the random source.  There are two main difficulties in this process. First of all,
systems deformed by an arbitrary source loose the spatial translational symmetry
and become difficult to handle, and secondly, taking the average is not trivial.
Two well-known methods to overcome both difficulties are the so-called ``replica
method'' and the ``Grassmann field method'' \cite{book:zee}.
Both aim at translating the said problem into an ordinary QFT calculation such that
more tools are at one's hand. The replica method involves duplicating the 
theory into multiple copies, introducing a mixing term that couples 
them, and taking the limit that the number of copies approaches zero.
While the replica method is an elegant formalism and a vast amount of literatures
are devoted to the topic, it is not always easily implemented in
many physically interesting systems, partially due to the bizarre limit one has to take.

Ever since its discovery \cite{Maldacena:1997re, Gubser:1998bc, Witten:1998qj}, 
AdS/CFT correspondence has been applied widely
to many types of systems. It did not take long before people attempted
to utilize its power to advance our understanding in condensed matter
theories (see, for example, \cite{Hartnoll:2009sz,Herzog:2009xv, Sachdev:2010ch, 
Bayntun:2010nx} and the references therein).  Given that fairly simple 
prescription for holographic duals for CFTs with multi-trace deformations 
was known \cite{Witten:2001ua}, for reasons to become clear below, 
it naturally leads to the proposal of implementing the replica 
method in holography and the hope for making progress in 
disorder problems \cite{Fujita:2008rs}.  Some preliminary attempts based 
on this idea to evaluate the DC conductivity have been made in~\cite{Hung:2010pe}.

Apart from all the technical subtleties, part of the weakness of
the discussions in \cite{Hung:2010pe} is that 
the bulk geometry considered there was pure AdS space, corresponding
to a system at zero temperature and with a zero charge density.
One would much prefer to consider models that are more realistic, 
in which both the temperature and the charge density are finite. Indeed, 
holographic models of such kind exist. One of the famous stories in this category is
the model for holographic superconductors where a hairy black hole replaces 
the pure AdS geometry \cite{Hartnoll:2008vx, Hartnoll:2008kx, Herzog:2010vz}. 
Chances appear to be much better that one finds more interesting results 
there since it is possible to have a non-zero background profile
for some bulk scalar field, and if one lets the random disorder
couple to the operator dual to that field, the background geometry
as well as all the perturbations above that are sensitive to the presence 
of disorder, leading to direct corrections to boundary-to-boundary correlation
functions.

In this note, we show that, despite the above naive expectation,
random disorder has zero corrections toward any connected correlation
functions in the leading order of the large-$N$ expansion, provided 
that the symmetry among replicas is not broken.  The last condition is 
a subtle assumption and systems known to violate it exist \cite{Binder:1986zz}.
Those must be studied in separation. It must be emphasized that
observables depending on the disorder in a non-trivial way certainly
exist (such as that in \cite{Hartnoll:2008hs}) 
and may be evaluated using the replica method. 
But the connected correlation functions themselves in the leading order
of $N$ is independent of the disorders that have a Gaussian distribution profile,
and any quantities determined by the connected
correlation functions through a linear relation, such as the DC 
conductivity, are not corrected either.
These results resemble very strongly the case where a single field 
is coupled to a random external source in a free QFT, even though
CFTs with semi-classical AdS duals are strongly coupled.  
We explain our statement using the replica method
both by the holographic duality as suggested in
\cite{Fujita:2008rs} and field theory analysis perturbatively 
in terms of the strength of random disorder for a toy model that
has a similar large-$N$ expansion. Both methods agree well.

At the sub-leading order, random disorder may lead to non-trivial
corrections. Based on our understanding gained from the field theory
model, we propose ways to evaluate those effects holographically.
Unlike the leading order effects, however, we are only able to 
do so perturbatively in terms of the strength of disorder. When the disorder
strength is normalized in the standard way as discussed in literatures,
higher order contributions are further suppressed by more powers of 
$N$ and become less and less important. In fact, it is not 
useful to evaluate them if one does not also include many other 
competing effects, such as the non-planar contributions at the same time.

We also discuss the possibility of making the disorder stronger
as $N\rightarrow \infty$ such that the non-vanishing corrections can
compete against other leading order physics in the large-$N$
expansion. In fact, it appears to be the most natural
and interesting choice since higher order contributions in terms of
the disorder strength are no longer suppressed and all appear at the 
leading order in the large-$N$ expansion.  We will show that this 
choice amounts to an unusual normalization for the double-trace 
deformation in a CFT. Such a normalization is usually considered pathological 
because the deformation destroys the large-$N$ expansion in ordinary CFTs.
But in the context of the replica method, the sickness is cured by
the appearance of a new small parameter, namely the number of the replicas.
In the queer limit that this number approaches $0$, a consistent 
large-N expansion may exist, leading to the hope that a new holographic
model can be defined and it captures the effects of disorder nonperturbatively.
Unfortunately, such a theory remains unknown to us.

The rest of the paper is organized as the following: in the next section,
we briefly explain the replica method emphasizing the ``magic limit'' in which
it ``converts'' disconnected correlation functions into properly normalized 
connected ones.  In Sec. \ref{sec:replica_ads} we implement
the replica method in holography as it is proposed in \cite{Fujita:2008rs}.
In Sec. \ref{sec:leading}, we present the leading corrections caused
by the disorder to the one- and two-point correlation functions using 
the AdS/CFT correspondence and show that they vanish when the symmetry
among the replicas are preserved.  In Sec. \ref{sec:matrix}, we reproduce
the same results perturbatively in the matrix model by field theory methods.
In Sec. \ref{sec:sub-leading}, we discuss how one may evaluate
the sub-leading non-zero effects, and the possibility of making them stronger and defining
a new type of holographic dual for CFTs with an unconventionally normalized
double-trace deformation.  We make some further discussions in Sec. \ref{sec:conclusion} 
and present a few more details in the appendix.

\section{The old holographic replica method and the correlation functions}
In this section, we briefly introduce the replica method and
provide an essential set of identities needed for later discussions.  We also
explain how one may implement the replica method in AdS/CFT as it
is discussed in literatures, where the disorder strength is normalized most conveniently
such that its leading contributions can be easily evaluated using known
methods. Unfortunately, as we show in the last part of this section, 
if one is only interested in connected correlation functions, such
contributions, even though being non-perturbative in terms of the
disorder strength, are simply zero, assuming that the symmetries among 
replicas is preserved.

\subsection{A brief introduction to disorder problem and the replica method}
\label{sec:replica}
The physics problem to consider
is evaluating the correlation functions in some QFT at the presence of a 
random external source \emph{averaged} over its probability distribution.
Let us denote the random source as $V(x)$ and couple it to an operator $O(x)$ in the QFT. 
For any fixed $V(x)$, the action for the QFT is deformed into
$S_V=S[X]+\int \ud^d x V(x) O(x)$, where $S[X]$ is the action for the original theory
and $X$ represents collectively all the fundamental fields.
As an example, the averaged expectation value for $O$ is given by
\[
\overline{\vev{O(x)}}=\int\uD[V] P[V]
\frac{\int \uD[X] O(x) e^{iS[X]+\int\ud^d x O(x) V(x)}}{
	\int \uD[X] e^{iS[X]+\int\ud^d x O(x) V(x)}}\,.
\]
Here, $P[V]$ is a predetermined probability distribution functional for $V(x)$.
In real materials, such a distribution functional is usually localized at
the time-independent configurations for $V(x)$, corresponding to impurities
that only break spatial homogeneity but preserve the energy conservation law. 
Throughout this discussion,
we would choose to ignore this subtlety and simply assume that $P[V]$ takes
the simplest possible Gaussian form as
\[
P[V]=e^{-\frac{1}{2f}\int\ud^d x V(x)^2},
\]
where we introduced the ``strength of the disorder'' $f$, assumed
to be a constant below.  
To make contact with more realistic physical systems, one should
promote $f$ to an operator, or in the momentum space, a general function of
the frequency and momentum. In particular, one may multiply $f$ by $\delta(\omega)$,
where $\omega=k^0$ is the frequency, so that $V(x)$ is restricted to be
time-independent. In the real-time formalism, this delta-function is
needed to recover unitarity.

As usual, we define the generating functional 
\[
Z_V[J]\equiv\int \uD[X] e^{S[X]+\int\ud^d x \left(O(x) V(x)+O(x) J(x)\right)}\,,
\]
and write 
\[
\bar{\vev{O(x)}}=\int \uD[V] e^{-\frac{1}{2f}\int\ud^d x V(x)^2}
\frac{\delta\ln Z_V[J]}{\delta J(x)}\,.
\]
The replica method is invented based on the following identity
\begin{equation}
\label{eq:master}
\lim_{n\rightarrow 0}\frac{1}{n} 
\frac{\delta^m Z[J]^n}{\delta J(x_1)\delta J(x_2)\dots\delta J(x_m)}
=\frac{\delta^m \ln Z[J]}{\delta J(x_1)\delta J(x_2)\dots\delta J(x_m)}\,.
\end{equation}
$Z[J]^n$ can be expressed explicitly by rewriting the same path-integral $n$ times:
\[
Z^n_V[J]\equiv\int \Pi_{i=1}^n\uD[X_i] e^{\sum_{i=1}^n S[X_i]
	+\int\ud^d x \sum_{i=1}^n O_i(x)\left(V(x)+J(x)\right)}\,.
\]
Here, we duplicated the same theory, including both the fields and
the action, for $n$ times, and hence the name of ``replica method''.  
While $n$ is defined as an integer in this procedure, we must assume 
that the physical observables can be understood as analytic functions 
of $n$ so that the limit $n\rightarrow 0$ is defined. 

Granted that the limit in Eq. \eqref{eq:master} makes sense, the magic of replica method
\begin{equation}\label{eq:magic}
\lim_{n\rightarrow 0} \frac{1}{n}
\vev{\sum_{i=1}^n O_i(x_1)\sum_{i=1}^n O_i(x_2)\dots\sum_{i=1}^n O_i(x_m)}^\circ
=\vev{O(x_1) O(x_2)\dots O(x_m)}^c
\end{equation}
follows immediately.  The left-hand side of this equation contains
the \emph{unnormalized disconnected} correlation
functions, denoted by the notation $\vev{\ast}^\circ$
throughout this manuscript,  evaluated in the theory with replicas,
and the right-hand side is the \emph{normalized totally connected} correlation
functions, denoted by the notation $\vev{\ast}^c$,
in the original single-copied theory. The former, if divided by
$n$, approaches the latter in the limit $n\rightarrow 0$.

Averaging both sides over the distribution $P[V]$, we find
\[
\bar{\vev{O(x_1) O(x_2)\dots O(x_m)}^c}=\lim_{n\rightarrow 0}\frac{1}{n}
\int \uD[V] e^{-\frac{1}{2f}\int\ud^d x V(x)^2}
\frac{1}{n}\frac{\delta^m Z^n_V[J]}{\delta J(x_1)\delta J(x_2)\dots \delta J(x_m)}\,.
\]
Assuming that one can interchange the order of the limit
$n\rightarrow 0$ and the functional integral for $V$,
we define
\[
Z^{(n)}_f[J]\equiv\int \Pi_{i=1}^n\uD[X_i] e^{\sum_{i=1}^n S[X_i]
	+\int\ud^d x \left(\delta\mathcal L+\sum_{i=1}^n O_i(x)J(x)\right)}\,,
\]
where 
\begin{equation}
\label{eq:double_deform}
\delta\mathcal L=\frac{f}{2}\left(\sum_i^n O_i\right)^2
\end{equation}
is the quadratic deformation, or in a CFT, the ``double-trace'' deformation,
and arrive at our master formula
\[
\begin{split}
\bar{\vev{O(x_1)O(x_2)\dots O(x_m)}^c}=&
\lim_{n\rightarrow 0}\frac{1}{n}
\frac{\delta^m Z^{(n)}_f[J]}{\delta J(x_1)\delta J(x_2)\dots \delta J(x_m)}\\
=&\lim_{n\rightarrow 0}\frac{1}{n}\vev{\sum_i^n O_i(x_1)
\sum_j^n O_j(x_2)\dots \sum_k^n O_k(x_m)}_f^\circ\,.
\end{split}
\]
We add a subscript ``$f$'' for the last correlation function in
the above equation to emphasize that it is evaluated
in the theory with replicas plus the quadratic 
deformation~\eqref{eq:double_deform}.

Particularly, for the vacuum expectation values, we have \footnote{The last step
assumed the symmetry among replicas are unbroken.}
\[
\bar{\vev{O}}=\lim_{n\rightarrow 0}\frac{1}{n}\vev{\sum_i^n O}^\circ_f
=\lim_{n\rightarrow 0}\vev{O_1}^\circ_f\,,
\]
and the connected two-point correlation functions
\begin{equation}\label{eq:2pt}
\overline{\vev{O(x_1) O(x_2)}^c}
=\lim_{n\rightarrow 0}\frac{1}{n}
\vev{\sum_{i=1}^n O_i(x_1) \sum_{j=1}^n O_j(x_2)}^\circ_f\,.
\end{equation}
It is easily verified that
\begin{equation}\label{eq:2pt_11_12}
\begin{split}
\overline{\vev{O(x_1) O(x_2)}}
=&\lim_{n\rightarrow 0}\vev{O_1(x_1) O_1(x_2)}_f^\circ\,,\\
\overline{\vev{O(x_1)}\vev{O(x_2)}}
=&\lim_{n\rightarrow 0}\vev{O_1(x_1) O_2(x_2)}^\circ_f\,.
\end{split}
\end{equation}
The particular values of the two subscripts for both operator $O_i$ 
on the right-hand side of both equations above are not important other than 
that they must equal for the first equation and different for the second.

It is sometimes useful to make the follow rotation among the replicas:
\begin{equation}\label{eq:tilde_basis}
\tilde O_i=\sum_j a_{ij} O_j\,,
\end{equation}
where $a_{nj}\equiv\frac{1}{\sqrt{n}}$ and
\[
a_{ij}\equiv\frac{1}{\sqrt{i(i+1)}}\left\{
\begin{array}{rr}
1,\quad& 1\le j\le i\\
-i,\quad& j=i+1\\
0,\quad & j>i+1
\end{array}\right.
\qquad\textrm{for~}i<n\,,
\]
easily verified to be orthogonal. Straightforward calculations show
\[
\begin{split}
\uvev{O_n O_n}=&\frac{1}{n}\left[\uvev{\tilde O_n \tilde O_n}
+(n-1)\uvev{\tilde O_{n-1}\tilde O_{n-1}}\right]\,,\\
\uvev{O_n O_{n-1}}=&\frac{1}{n}\left[\uvev{\tilde O_n \tilde O_n}
-\uvev{\tilde O_{n-1}\tilde O_{n-1}}\right]\,.
\end{split}
\]
By Eq. \eqref{eq:2pt_11_12}, in the limit $n\rightarrow 0$, we have
\begin{equation}\label{eq:2pt_tilde}
\begin{split}
&\bar{\vev{O O}}=\lim_{n\rightarrow 0}
\uvev{O_n O_n}=\lim_{n\rightarrow 0}\uvev{\tilde O_{n-1}\tilde O_{n-1}}
+\lim_{n\rightarrow 0}\frac{1}{n}\left[\uvev{\tilde O_n \tilde O_n}
-\uvev{\tilde O_{n-1}\tilde O_{n-1}}\right]\,,\\
&\bar{\vev{O}\vev{O}}=\lim_{n\rightarrow 0}
\uvev{O_n O_{n-1}}=\lim_{n\rightarrow 0}\frac{1}{n}\left[\uvev{\tilde O_n \tilde O_n}
-\uvev{\tilde O_{n-1}\tilde O_{n-1}}\right]\,.
\end{split}
\end{equation}
Therefore, we also find 
$\bar{\vev{O O}^c}=\lim_{n\rightarrow 0}\uvev{\tilde O_{n-1}\tilde O_{n-1}}$.
Comparing this to Eq. \eqref{eq:2pt} implies the replica
method is only self-consistent if the following two limits coincide:
\[
\lim_{n\rightarrow 0}\uvev{\tilde O_{n}\tilde O_{n}}
=\lim_{n\rightarrow 0}\uvev{\tilde O_{n-1}\tilde O_{n-1}}\,.
\]
In fact, it is required that the difference between
$\vev{\tilde O_n\tilde O_n}^\circ$ and $\vev{\tilde O_{n-1}\tilde O_{n-1}}^\circ$
is $O(n)$ so that $\bar{\vev{O}\vev{O}}$ would not diverge, 
as shown in Eq. \eqref{eq:2pt_tilde}.

We mention that rotating into the basis of $\tilde O_i$ is not extremely useful in general
since both the action $S$ and the operator $O[X]$ are typically nonlinear
functionals of $X$, and not only it is non-trivial to implement such a 
rotation in terms of the fundamental fields $X_i$ but also it
creates more mixing in other parts of the action leading to no simplifications.
Only in the trivial case explained in the Appendix \ref{app:free}, it is helpful simply because
the action is quadratic. It turns out, to the leading order of
the large-$N$ expansion, a class of problems that can be treated by holographic
methods discussed below appears to present another 
example where such a rotation is useful, leading to just as trivial 
results.

\subsection{The replica method in AdS/CFT}
\label{sec:replica_ads}
While the replica method is a beautiful idea, it is difficult to 
implement in a generic QFT, particularly when it is strongly coupled.
On the other hand, to the leading order in the large-$N$
expansion, holographic dual for a CFT deformed by a multi-trace term was
known \cite{Witten:2001ua} and a vast amount of literatures were devoted
to studying those theories \cite{Aharony:2001pa, Berkooz:2002ug, Sever:2002fk,
Aharony:2005sh, Elitzur:2005kz, Hartman:2006dy, Papadimitriou:2007sj,Compere:2008us,
Vecchi:2010jz}.  One naturally hopes that implementing the replica method in the 
AdS/CFT correspondence may be tractable and bring us to better
understandings to disorder problems.

First, let us set the number of the replicas to $1$ and briefly mention how 
the multi-trace deformation of the form
\begin{equation}\label{eq:multi-trace}
\delta \mathcal L=W[O]
\end{equation}
is treated holographically. Here $O$ is some single-trace operator
in the CFT and $W[x]$ is a polynomial of $x$. 

The bulk geometry we have in mind is an $d+1$ dimensional asymptotically
AdS space.  It may be an AdS charged black hole, or an hairy charged black 
hole as discussed in \cite{Hartnoll:2008vx, Hartnoll:2008kx, Herzog:2010vz}. 
We will not need to specify the geometry in details in this discussion.

Let $\psi$ be the bulk field dual to $O$.  in such an asymptotic AdS space.
Near the boundary, if the metric approaches the form 
$\ud s^2=z^{-2}(\ud z^2+\ud x^\mu \ud x_\mu)$, $\psi_i$ behave as
\[
\psi\sim \alpha z^{\Delta_+}+\beta z^{\Delta_-}\,,
\]
where $\Delta_\pm=d/2\pm\sqrt{d^2/4+m^2}$.
We identify $\Delta_-$ with the conformal dimension of $O$.
By the holographic principle, when $\psi$ satisfies the boundary condition
$\alpha=0$ near $z=0$, $\beta$ is identified with $\vev{O}$ in a CFT without deformations.
When the CFT is deformed by a multi-trace term~\eqref{eq:multi-trace},
it was proposed that, to the leading order of $N$, 
$\psi$ must satisfy the modified boundary condition \cite{Witten:2001ua}
\[
\alpha=\frac{\delta W[\beta]}{\delta \beta}\,,
\]
and the rest of the AdS/CFT dictionary remains unchanged. 
The double-trace deformation usually breaks the conformality
and lead to non-trivial RG flow \cite{Witten:2001ua, Vecchi:2010jz}. 
We have above restricted ourselves
in the window such that the deformation is relevant, i.e. $2\Delta_O=2\Delta_-<d$.
Correspondingly, the mass for the bulk scalar must be within the
window $-d^2/4<m^2<-d^2/4+1$ so that alternative AdS quantizations
are allowed \cite{Klebanov:1999tb}

Given this simple prescription, it is suggested 
that the replica method may be implemented in the AdS/CFT correspondence
as the following \cite{Fujita:2008rs}. 
Duplicate all the fields in the bulk, including the metrics,
by $n$ times and set the boundary condition for the fields dual to
the operator $O$ as mentioned above.  For example, consider a phenomenological 
model given by the Lagrangian
\begin{equation}\label{eq:ads_replica}
\begin{split}
\mathcal L=\sum_i^n\big[&R(g_i)+6-\frac{1}{4} F_{\mu\nu} F^{\mu\nu}(A_i)
-(\partial_\mu \psi_i-iqA_\mu \psi_i)
	(\partial^\mu\psi_i^\dag+iq A^\mu \psi_i^\dag)\\
&\;-m^2\psi_i\psi^\dag_i
	-\lambda |\psi_i|^4+\dots
	\big]\,,
\end{split}
\end{equation}
where we have set the bulk curvature and the Planck mass to $1$,
and introduced $n$ copies for every field, including the metrics 
$g_i$, the gauge fields $A_i=A^i_\mu\ud x^\mu$, and the complex scalars
$\psi_i$. The scalar $\psi$ would be dual to
the operator $O$ in the CFT that is coupled to the 
impurity~\footnote{Here, the operator $O$ is charged, and so must
be $V$ that's coupled to it. The double trace deformation term would
then become $\frac{f}{2}\left(\sum_i O_i\right)^\dag \left(\sum_j O_j\right)$.
Similarly, the correlation functions to evaluate should 
be $\vev{O^\dag O}^c$. For the simplicity of notations, we would 
ignore the complex conjugation though.}.  
Toward the boundary, $\psi_i\sim \alpha_i z^{\Delta_+}+\beta_i z^{\Delta_-}$,
for which we should set the boundary conditions as
\begin{equation}\label{eq:mixing_bc}
\alpha_i=f \sum_i^n \beta_i\,.
\end{equation}
Such boundary conditions not only mix different copies of the scalar fields,
but, through loop corrections, also mix all the gauge and gravitational fields,
making all but one linear combination of the replicas massive, reflecting
the fact that only one conserved energy-momentum tensor and $U(1)$ current
are preserved on the CFT side due to the mixing double-trace deformation.
We will evaluate the correlation functions in this 
theory by the standard AdS/CFT duality and take the limit 
$n\rightarrow 0$.

We should mention discussions on applying
holographic models for CFTs with double-trace deformations to condensed
matter problems exist in literatures, and they are not always
motivated by the replica method explained above (see, for example,
\cite{Faulkner:2010gj}).  When the multi-trace deformations
are introduced for other reasons, one is probably not interested in taking
the limit $n\rightarrow 0$, or such a limit simply does not exist,
in which case the following discussions of course do not apply.

\subsection{The vanishing leading order corrections}
\label{sec:leading}
Let us first examine the possible corrections to the 
vacuum expectation values. For this purpose,
we look for the background solutions to the equations of motion 
led by the Lagrangian \eqref{eq:ads_replica}. 
Schematically, we can write the equations as
\[
\mathbf\Psi_i[\psi^{cl}_i, g^{cl}_i, A^{cl}_i]=0,\quad
\mathbf E_i[\psi^{cl}_i, g^{cl}_i, A^{cl}_i]=0,\quad
\mathbf M_i[\psi^{cl}_i, g^{cl}_i, A^{cl}_i]=0,\quad
i=1,2, \dots, n\,,
\]
where $\mathbf\Psi$, $\mathbf E$, and $\mathbf M$ represent 
the Klein-Gordon equations, the Einstein's equations, and
the Maxwell's equations respectively.  The solutions for the metrics
must be asymptotically AdS but may develop a black hole in the bulk.
These equations are nonlinear and known analytic solutions
are rare, but can often be solved numerically.
These equations do not mix fields for different replicas,
but the boundary conditions \eqref{eq:mixing_bc} do.

We assume that the background solutions respect the
translational symmetry in the $d$ flat dimensions
and are $x^\mu$-independent. Given such assumption, 
all the equations above form a system of ordinary 
differential equations and fields in different replica share
identical equations as well as boundary conditions. If the vacuum is unique, 
we must find the solutions for all replicas are identical.
In the scenario that there exist multiple vacua in the bulk,
the solution that corresponds to the least action should 
be favored and the action in the bulk is simply a sum of
those of all replicas, it is natural to assume
that each replica picks the same solution. However, we should emphasize
that there exist possible scenarios where this symmetry is broken
if more complicated potentials are included, and
when such phenomenon happens, the problems is substantially
more complicated \cite{Fujita:2008rs, Binder:1986zz}.  We will not further discuss
this possibility throughout this paper. Given these assumptions, 
mixing boundary conditions \eqref{eq:mixing_bc} automatically
decouple and become equivalent to
\begin{equation}\label{eq:bc_background}
\alpha_i=nf\beta_i\,,\quad i=1, 2, \dots, n\,.
\end{equation}

It is manifest from Eq. \eqref{eq:bc_background} that the background profiles
depend on the strength of the double trace deformation $f$ only through the 
combination of $\tilde f\equiv nf$. Hence, taking $n\rightarrow 0$ limit becomes equivalent
to turning off the disorder. For example, $\bar{\vev{O}}$ is given by
\[
\bar{\vev{O}}=\lim_{n\rightarrow 0}\frac{1}{n}\vev{\sum_{i=1}^n O_i}_f
=\lim_{n\rightarrow 0}\vev{O_1}_f=\lim_{n\rightarrow 0}\beta_1\big|_{\alpha_1=n f\beta_1}\,.
\]
The limit $n\rightarrow 0$ and $f\rightarrow 0$ is indistinguishable 
in the last expression and it is necessarily true that
\[
\bar{\vev{O(x)}}_f=\vev{O(x)}_{f=0}\,,
\]
as if there is no random disorder at all.
Here, a key requirement is that the background solutions 
$(\psi^{cl}, g^{cl}, A^{cl})$ vary with respect to
$\tilde f$ smoothly as $\tilde f$ approaches $0$.
It certainly holds true in the present case with the assumption
that the solutions for all replica are identical, and in the
limit $\tilde f\rightarrow 0$, the standard boundary condition is recovered 
from Eq. \eqref{eq:bc_background}.

Now, let us turn to the two-point correlation functions. To this end, we should
consider the fluctuations above the background profiles:
$g_i\rightarrow g^{cl}_i+\delta g_i$, 
$A_i\rightarrow A^{cl}_i+\delta A_i$, 
and 
$\psi_i\rightarrow \psi^{cl}_i+\delta \psi_i$, and solve for them
by perturbing the equations of motions to the first order.
Schematically, we can write:
\begin{equation}\label{eq:pert}
\hat K[g^{cl}, A^{cl}, \psi^{cl}] \delta X_i=0\,,
\end{equation}
where $\delta X$ denotes the perturbations $(\delta g, \delta A, \delta\psi)$
collectively, which may be thought of a column vector and $\hat K$ 
is a linear second order differential operator obtained by expanding 
the Einstein's equation, the Maxwell equation and the 
Klein-Gordon equation to the first order of the perturbations above
the background $(g^\textrm{cl}, A^\textrm{cl}, \psi^\textrm{cl})$.
Fields in different replicas must all be allowed to fluctuate freely
and their perturbations are generically different.
However, here we have the extra advantage that the perturbed equations Eq.~\eqref{eq:pert}
form a linear system.  We may apply the rotation 
introduced in Sec. \ref{sec:replica}, quite similar to the 
way we solve the free theories outlined in Appendix \ref{app:free}. 
Define,  for each $X_i\in \{g_i, A_i, \psi_i\}$,
\[
\tilde X_i=\sum_{j=1}^n a_{ij} X_j\,,
\]
where $a_{ij}$ are given in Sec. \ref{sec:replica}.  According
to Eq.~\eqref{eq:2pt} and the AdS/CFT correspondence, we find
\begin{equation}\label{eq:2pt_ads}
\bar{\vev{O_X(x) O_X(y)}^c_{\textrm{CFT}}}=
\lim_{n\rightarrow 0} 
\vev{\tilde O_{X_n}(x)\tilde O_{X_n} (y)}^\circ
=\lim_{n\rightarrow 0}
\Pi_{\tilde X_n \tilde X_n}(x, y)\,,
\end{equation}
where $\Pi_{\tilde X_n\tilde X_n}$ is the boundary-to-boundary
propagator for the field $\tilde X_n$ and
$O_X$ denotes the operator that is dual to $X$. Some objections may be
raised against the last step above. In the intermediate step,
one is supposed to evaluate the unnormalized disconnected two point function
for the operator $\tilde O_{X_n}$, but by the AdS/CFT dictionary, the 
boundary-to-boundary propagator is identified with the normalized
connected correlation function. In this context, the difference is
not important. First of all, the normalization is irrelevant since 
a key assumption for replica method to work is that $Z_n\rightarrow 1$ in
the limit $n\rightarrow 0$, where $Z_n$ is the partition function.
While this limit is subtle as $Z_n$ is usually divergent and one must
properly regularize it, provided the proper regularization is found, 
we may interchange the normalized and unnormalized correlation 
functions in the limit $n\rightarrow 0$. Secondly, the disconnected
correlation function differs from the connected one by a possibly nonzero
term $\vev{\tilde O_{X_n}}^2$. But this term is always proportional
to $n$, as is easily verified, and vanishes in the limit $n\rightarrow 0$.
A quick way to see this must be true is to recall that
\[
\lim_{n\rightarrow 0}\vev{\tilde O_{X_n}\tilde O_{X_n}}=
\lim_{n\rightarrow 0}\vev{\tilde O_{X_{n-1}}\tilde O_{X_{n-1}}}
\]
is essential in replica method, and by construction $\vev{\tilde O_{X_{n-1}}}=0$
\footnote{Once again, the symmetry among all replicas is used}.
Therefore, we may as well replace the field $\tilde X_n$ by $\tilde X_{n-1}$,
which leads us to the conclusion below more quickly.

In any case, the boundary-to-boundary propagator 
for $\tilde X_n$ is given by the solution for $\delta\tilde X_n$. In 
the tilde-basis, the boundary condition is ``unmixed''. Depending
on the correlation functions one is interested in, boundary conditions
for fields other than $\tilde\psi_n$ are standard ones and only that for 
$\delta\tilde\psi_n$ is modified. For example, if one
wish to evaluate $\vev{\tilde O_{X_n}(x)\tilde O_{X_n}(y)}^c$, corresponding to the response
of the system with respect to an infinitesimal change of the source that's coupled
to $O$, one may set
\[
\delta\tilde\alpha_n=1+nf\delta\tilde\beta_n\,, 
\]
for $\delta\tilde\psi_n$, and make sure other fluctuations vanish near the AdS boundary, 
after proper regularizations if necessary.  The averaged two-point correlation function
for the order parameter, $\bar{\vev{O(x) O(y)}^c}$,
is then given by $\lim_{n\rightarrow 0}\delta\tilde\beta_n$.
Both the boundary condition and the operator $\hat K$
depend on $f$, and $\hat K$ acquires its dependence through
the background solutions. However, the appearance of $f$ here in the boundary condition
is again in the combination of $\tilde f=nf$, 
and taking the limit $n\rightarrow0$ is still equivalent
to taking the limit $f\rightarrow 0$. The same is true for
$\hat K$ simply because it happens to the background solutions 
as discussed. Hence, taking the limit $n\rightarrow 0$
completely eliminates the $f$-dependence in Eq.~\eqref{eq:2pt_ads}.

In fact, as is alluded to already, the consistency of the replica method demands 
that the same limit is achieved by considering $\delta\tilde\beta_{n-1}$ instead,
for which the standard boundary condition applies. 
The only $f$-dependence of $\delta\tilde\beta_{n-1}$
comes through $\hat K$. We, therefore, are led to the conclusion even faster 
that the connected two-point correlation function are insensitive to the 
presence of the disorder, when the double-trace deformations is taken into account
by the generalized boundary conditions as explained and with the assumption that the symmetry
among the replicas is unbroken.  Despite that it may appear counter-intuitive,
this conclusion is explicitly verified by the results presented in
\cite{Fujita:2008rs}.

We should mention that it is certainly not true that no observables
depend on $f$ exist. For example, both
\[
\begin{split}
\bar{\vev{O(x)O(y)}_{\textrm{CFT}}}
=&\lim_{n\rightarrow 0}\frac{1}{n}\left[
\delta\tilde\beta_n+(n-1)\delta\tilde\beta_{n-1}\right]\,,\\
\bar{\vev{O(x)}_{\textrm{CFT}}{O(y)}_{\textrm{CFT}}}
=&\lim_{n\rightarrow 0}\frac{1}{n}\left[
\delta\tilde\beta_n-\delta\tilde\beta_{n-1}\right]\,\\
\end{split}
\]
depend on $f$ non-trivially because the difference
between $\delta\tilde\beta_n$ and $\delta\tilde\beta_{n-1}$ is $O(n)$.

Similarly, to evaluate the current-current correlation function, we should solve
for $\delta\tilde A_n$ with the boundary conditions
\[
\delta\tilde A_n=1\,,\qquad \delta\tilde \alpha_n=nf\delta\tilde\beta_n\,,
\]
and all other perturbations approach $0$ near the AdS boundary.
By the same reasoning, in the limit $n\rightarrow 0$,
the averaged two-point function should just become trivial:
\[
\bar{\vev{J_\mu(x) J_\nu(y)}^c}=\vev{J_\mu(x) J_\nu(y)}^c_{f=0}\,,
\]
as if $f=0$ in the first place.

So, given the said assumptions, we find disorder has no effect on the 
averaged connected correlation functions in the leading order of the large-$N$ expansion
in any CFT that has a holographic dual described by~\eqref{eq:ads_replica}.
The way we arrive at this conclusion is
very similar to how we solve the same problem in a free theory deformed by
disorder as shown in Appendix \ref{app:free}. Certainly, the symmetry among
the replicas play an essential role which allows one to diagonalize the
system in its holographic description very easily.

\section{A toy field theory model}
\label{sec:matrix}
What we found above is not immediately intuitive considering the
CFT is supposedly strongly coupled.  We seek a field theory understanding
toward the same problem in this section.  Besides the very much needed
reassurance to verify that the holographic principle does not break down,
a field theory understanding might also allow us to probe further 
the sub-leading corrections, if any, not captured by the
treatment presented above.

It is not easy to reproduce the results given above in an arbitrary CFT 
deformed by the disorder.  But, perturbatively in terms of the disorder 
strength $f$, we may make some attempts in a toy model that
share some important properties with theories that have a holographic dual.
We emphasize that when we claim to analyze the system perturbatively, we always
mean that we do so perturbatively with respect to the disorder 
strength $f$, and make no assumption about
the theory itself being strongly or weakly coupled or whether the theory
can be treated perturbatively with respect to any other couplings. 
As long as we can tune the parameter
$f$ to be sufficiently small, we assume that a perturbative expansion
for the double-trace deformation, or more generally the multi-trace deformation,
in the orders of $f$ can be defined even if the theory is strongly coupled.

The toy model we present here is the matrix theory. While it is not a CFT,
we may proceed since the discussions below solely rely on the assumptions that the
disorder can be treated perturbatively and the theory has a large-$N$ expansion 
and well-defined t' Hooft limit in which correlation functions 
have the similar factorization properties as those in the matrix model.

The theory is described by a Lagrangian:
\begin{equation}\label{eq:matrix_L}
\mathcal L=\frac{1}{g^2}\Tr\left[
\partial M\partial M+ a M^2 + b M^4+\dots. \right]\,,
\end{equation}
where $M$ is a $N\times N$ Hermitian matrix. Only the
single-trace terms are present in the unperturbed theory.  We use
the standard double-line notations for the propagators and vertices.
In the t' Hooft limit, $g^2 N\rightarrow \lambda$, all connected
correlation functions are of the order $N^2$ with planar diagrams
being the leading contributions. Non-planar diagrams are suppressed 
by additional factors of $N$.  This is easily
understood if one recall that each vertex contributes a factor of
$g^{-2}$, each propagator contributes a factor of $g^2$, and each closed
loop contributes a factor of $N$. Every connected diagram carries
an overall factor $N^\chi (g^2 N)^\#$, where $\chi$ is the Euler number 
for the surface on which the diagram can be drawn. Disconnected diagrams
that contain multiple connected components carry a factor of
$N$ that is the product of the factors of $N$ for every
individual connected piece, and therefore, is larger than $N^2$ generally.

Without introducing the replicas, let us first briefly explain what 
happens if the above theory is deformed by a multi-trace term
\begin{equation}\label{eq:matrix_model_multi_def}
\delta \mathcal L=\frac{f}{2} W[O], \qquad W[O]=N^{2-m} O^m\,.
\end{equation}
In particular, the double-trace 
deformation is given by the choice of $W[x]=x^2/2$.
We have chosen to normalize the singe-trace operator $O$ slightly unconventionally
but compensated for it by inserting the additional factor $N^{2-m}$ in
the multi-trace term so the theory is the same to those considered in 
\cite{Witten:2001ua}. The motivation for this choice of normalization is
that it reads $f O^2/2$ for quadratic deformation, corresponding to
a direct coupling between the disorder $V$ and the operator
$O$, when interpreted in the context of the replica method. 
We may choose to couple $V$ to some single-trace operator with a different normalization,
leading to weaker or stronger effect as we will discuss later.

\begin{figure}[ht!]
\begin{center}
\includegraphics[width=0.45 \textwidth]{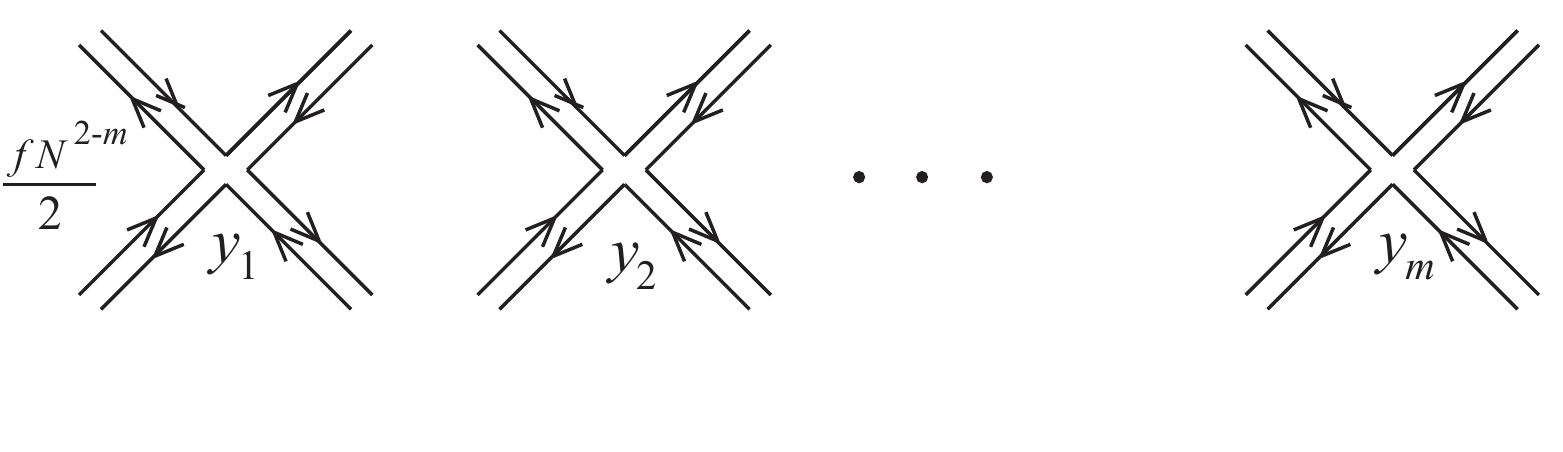}
\caption{\label{fig:split-vertices}Momentarily, think of the multi-trace vertex as 
a group of single-trace ones coincide.}
\end{center}
\end{figure}

To help with our counting of the powers of $N$ when multi-trace vertices
are present, we introduce a small trick. Let us split the multi-trace vertex
slightly so that it appears as separate single-trace ones,
as illustrated in Fig. \ref{fig:split-vertices}, and think of the original
vertex as the coincide limit of them, i.e. we would 
take the limit $y_i\rightarrow y$, $i=1, 2, \dots m$,
as shown in Fig.~\ref{fig:split-vertices}, so
they ``recombine'' into a single multi-trace vertex.
This helps because when the multi-trace vertices are involved, strictly
speaking one may no longer talk about planar diagrams as
the vertices themselves can not be drawn on a single plane. Once we
split them into multiple single-trace ones, the more familiar
counting methods can be applied, with only the caution that 
the group of vertices obtained by splitting a multi-trace one do not 
carry the usual factor of $g^{-2}$'s and one must be careful with how they
are normalized.  

Admittedly, taking the coincide limit of several single-trace 
vertices to form a multi-trace one is not trivial 
(if one actually does it literally), since various divergences
would arise in the coincide limit as usual. 
However, at least within this section, such a splitting and 
recombining process is merely a bookkeeping tool to help us with the counting.
One could, in principle, rephrase everything below without making
the reference to this trick at all, but the words become more clumsy as one
starts to implement the replica method.  Having said that, We assume
a properly regularization scheme can be applied
so that the limit-taking process actually makes sense. 
Since the regularization scheme itself does not involve the overall factors of $N$,
the counting below is not affected.  

\begin{figure}[ht!]
\begin{center}
\includegraphics[width=0.5 \textwidth]{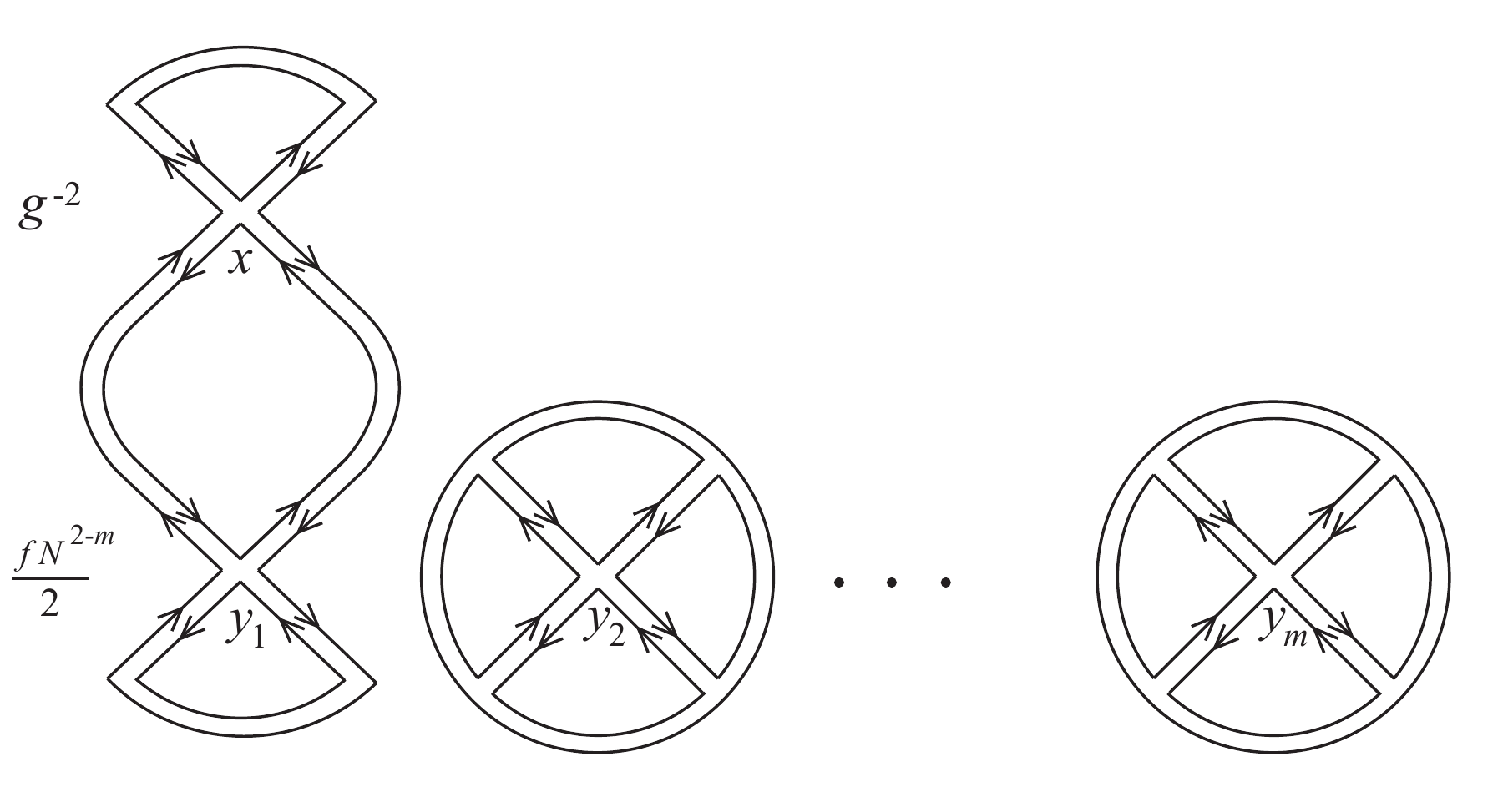}
\caption{\label{fig:witten-leading}The leading corrections to
$\vev{O}$ when $n=1$.}
\end{center}
\end{figure}

Let us consider the correction to the $\vev{O}$ due to 
a deformation term \eqref{eq:matrix_model_multi_def} in this theory.
We treat the deformation perturbatively, so we expand
$\exp\{\delta\mathcal L\}$ order by order in $f$, insert each term 
in the path-integral, and evaluate the path-integral
with the Lagrangian Eq.~\eqref{eq:matrix_L} without deformations.
To the leading order of $f$, the correction to $\vev{O}$ is
given by the integral
\begin{equation}\label{eq:deltaO}
\begin{split}
&\delta\vev{O(x)}\\
&=\frac{fN^{2-m}}{2}\int\ud^d y\, 
\vev{O(x) O(y)^m}^c=
\int\ud^d y\,\left[\vev{O(x) O(y)^m}-\vev{O(x)}\vev{O(y)^m}\right]\\
&=\frac{fN^{2-m}}{2}
\int\ud^d y\, \lim_{y_i\rightarrow y}\left[\vev{O(x) O(y_1) O(y_2)\dots O(y_m)}
-\vev{O(x)}\vev{O(y_1)O(y_2)\dots O(y_m}\right]\,.
\end{split}
\end{equation}
In the last step, the limit $y_i\rightarrow y$ is just a formal notation,
but, with those vertices separated, we could visualize this expression
more easily as Feynman diagrams that connect the vertex at $x$ to at least one of 
those at $y_i$, since diagrams that does not connect $x$ to any 
of the $y_i$'s are subtracted out. 
Notice that this is the only requirement for the above expression to be
non-vanishing and no more vertices are obligated to get connected.
Given the constraint, the more connected components there are, 
the more powers of $N$ this correlation function obtains. So, 
the most dominant diagrams that contribute to $\delta\vev{O}$
must be those that connect only $x$ to one of the $y_i$'s, but leaving all
the rest disconnected, as illustrated in Fig.~\ref{fig:witten-leading}.
Should all the vertices at $y_i$'s be the ordinary single-trace
ones, this diagram would be proportional to $N^{2m}$ at most.
But here, since the group of $y_i$'s together contributes
a factor of $f N^{2-m}$ in stead of $N^m$, such diagrams are of the order
$f N^{2m-m+(2-m)}=f N^2$, exactly the same as the usual planar ones.

We can express the diagrams in Fig. \ref{fig:witten-leading},
after some proper regularizations, as
\[
\delta\vev{O}=\frac{m f}{2} \int\ud^d y \vev{O(x) O(y)}^c \vev{O}^{m-1}
\]
where the extra factor of $m$ arises due to the fact
that one may choose to connect the vertex at $x$ to any one of the 
vertices at $y_i$ leading to equivalent terms.  If we translate this object by the 
AdS/CFT dictionary, after proper normalizations, this is the same as
\[
\delta\vev{O}=m \int\ud^d y\, \frac{\delta\beta(x)}{\delta\alpha(y)}\beta(y)^{m-1}
\]
since $\delta\beta(x)/\delta\alpha(y)$ is identified with
the two-point correlation function $\vev{O(x)O(y)}^c$ and $\beta=\vev{O}$.
At the same time, the right-hand side of this equation
is precisely the change of $\beta$, to the leading order of $f$, due to the change of
the boundary condition from $\alpha=0$ to $\alpha=mf\beta^{m-1}$, because 
$\delta\beta(x)=\int\ud^d y \frac{\delta\beta(x)}{\delta\alpha(y)}\delta\alpha(y)$,
and $\delta\alpha(y)=mf\beta(y)^{m-1}$.
The analysis generalizes to every higher order corrections
similarly, reproducing exactly the same results in the leading
order of $N$ as those given by the altered boundary condition
prescription in the holographic method.
While the analysis here is perturbative in terms of $f$, 
the same can be derived non-perturbatively by a saddle point calculation 
\cite{Witten:2001ua}.

We are now at the position to reproduce what has been found in the previous
section by implementing the replica method in the matrix model, and examine
what happens to the class of diagrams illustrated in Fig.~\ref{fig:witten-leading}
and why their effects disappear in the limit $n\rightarrow 0$.

Let us introduce the replicas so the Lagrangian~\eqref{eq:matrix_L} 
is duplicated by $n$ times, and consider order by order the correction 
to the correlation functions due to the multi-trace deformation 
\begin{equation}\label{eq:mixing-double-trace}
\delta\mathcal L=\frac{f}{2} W\left[\sum_i^n O_i\right]\,.
\end{equation}
In the replica method, $W[x]=x^2/2$. 

Consider $\delta\bar{\vev{O}}$, which, to the first order of $f$, 
is given by the integral 
\[
\begin{split}
\delta\bar{\vev{O(x)}}
=&\frac{f}{2}\lim_{n\rightarrow 0}\frac{1}{n}\int\ud^d y\,
\lim_{ y_{_{1,2}}\rightarrow y}
\vev{\sum_i O_i(x)\sum_j O_i(y_1)\sum_k O_k(y_2)}^\circ\\
=&\frac{f}{2}\lim_{n\rightarrow 0}\int\ud^d y\,
\lim_{ y_{_{1,2}}\rightarrow y}
\vev{O_1(x)\sum_j O_i(y_1)\sum_k O_k(y_2)}^\circ\,.
\end{split}
\]
We remind the readers that the correlation functions on the right-hand side
are evaluated in the theory whose Lagrangian is simply $n$ copies
of Eq. \eqref{eq:matrix_L} without the double-trace 
deformation~\eqref{eq:mixing-double-trace}. Before taking the limit
$n\rightarrow 0$, the disconnected correlation function in the last
line of the above equation contains several categories of contributions. 
Feynman diagrams that leave all $x$, $y_1$, and $y_2$ disconnected would acquire a factor
of $n^2$, diagrams similar to those in Fig.~\ref{fig:witten-leading}
that connect only $x$ to one of the $y_1$ or $y_2$ acquire a factor
of $n$, and lastly, diagrams that connect the vertices at all
three locations does not acquire any factor of $n$. This is shown
expicitly in Appendix~\ref{app:magic}.  So, in the limit 
$n\rightarrow 0$, only the diagrams in the last category survive.

\begin{figure}[ht!]
\begin{center}
\includegraphics[width=0.5 \textwidth]{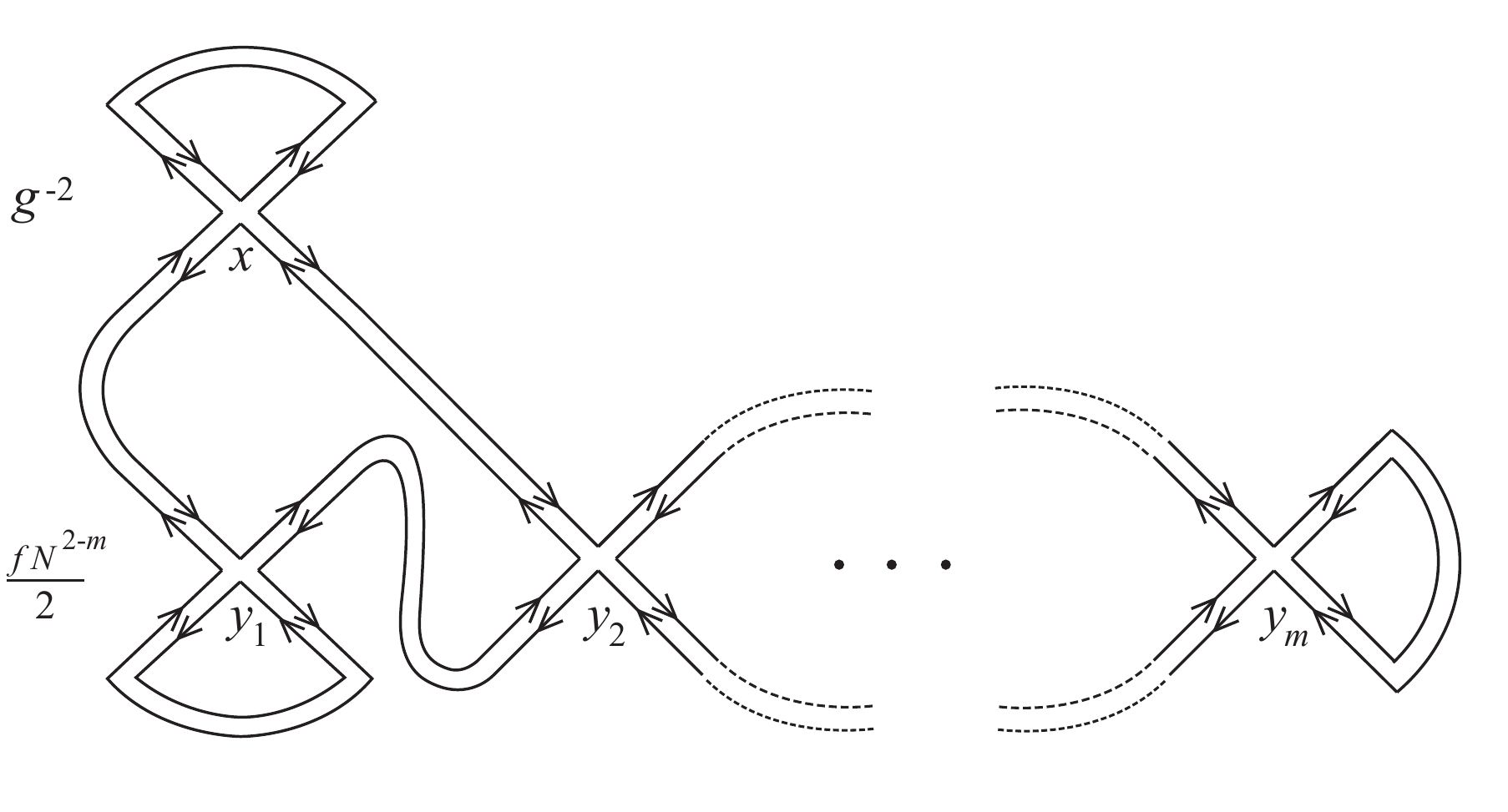}
\caption{\label{fig:replica-leading}The leading corrections to
$\bar{\vev{O}}$ when $n\rightarrow 0$.}
\end{center}
\end{figure}

There is an elegant way to make this discussion clearer and more 
readily generalizable to higher order corrections. Let us consider 
the following object:
\[
\mathcal C
=\lim_{n\rightarrow 0}\frac{1}{n}\int\ud^d y\,
\lim_{y_i\rightarrow y}
\vev{\sum_i O_i(x)\sum_j O_i(y_1)\sum_k O_k(y_2)\dots\sum_l O_l(y_m)}^\circ\,.
\]
As explained in Sec. \ref{sec:replica}, the power of
the replica method is that in the limit $n\rightarrow 0$,
it turns the above disconnected correlation function into
a normalized and fully connected one, i.e. by the identity~\eqref{eq:magic},
we find
\begin{equation}\label{eq:C_formula}
\begin{split}
&\mathcal C=\lim_{n\rightarrow 0}\frac{1}{n}\int\ud^d y\,
\lim_{y_i\rightarrow y}
\vev{\sum_i O_i(x)\sum_j O_i(y_1)\sum_k O_k(y_2)\dots\sum_l O_l(y_m)}\\
&\qquad\qquad=\int\ud^d y\lim_{y_i\rightarrow y}
\vev{O(x)O(y_1) O(y_2)\dots O(y_m)}^c\,.
\end{split}
\end{equation}
Here, we arrive at the essential difference between $\mathcal C$ 
and Eq.~\eqref{eq:deltaO}. The correlation function in the last expression above
is \emph{fully connected before the limit $y_i\rightarrow y$}.
One may visualize it as Feynman diagrams that contain a single connected 
component even when the multi-trace vertex is split.
We arrived at this formula quickly by citing the general results~\eqref{eq:magic}.  
There, it seems essential that all vertices
are located at separate positions and the validity of its application here might appear
questionable. An explicit calculation that makes no reference to the splitting 
and recombining process is presented in Appendix \ref{app:magic}, 
which both verifies the above formula for the case of a double-trace insertion
and serves to clarify what we mean exactly by ``the fully connected diagrams'' here.
We remind the readers that in the limit $y_i\rightarrow y$,
diagrams shown in Fig.~\ref{fig:witten-leading} are \emph{connected} too,
but become disconnected once the multi-trace vertex is split. The vitally 
important difference between those and the correlation functions
appearing in the last line in~\eqref{eq:C_formula} is that the latter
stays \emph{fully connected even} when the multi-trace vertices at 
$y$ is split into separated single-trace ones! We refer those them
as the ``super connected'' diagrams, as illustrated in Fig.~\ref{fig:replica-leading}.

Should the vertices at $y_i$'s be ordinary
single-trace ones, the ``super connected'' diagrams
would be $O(N^2)$ because one can draw a planar diagram that connects
all vertices. Here, however, the counting is somewhat different since
the group of vertices at $y_i$ contributes less powers of $N$.
We already know that diagrams illustrated in Fig.~\ref{fig:witten-leading}
are $O(N^2)$, which immediately implies that ``super connected'
diagrams being considered here are 
sub-dominant simply because they contain fewer number of connected components.
Indeed, in the case of a double-trace insertion, a quick counting
shows that the leading contributions for $\delta\bar{\vev{O}}$
is of the order $N^0$. Similar analysis may be carried out to higher
order corrections in terms of $f$ as well to multi-point correlation functions.

At this point, it is not surprising at all
that if one follows the standard holographic prescription for
CFTs with double-trace deformations, which
takes into account only its leading effects at the order of $N^2$,
one finds all connected correlation functions are not affected
by disorders once the limit $n\rightarrow 0$ is taken. In other words, 
in any CFT that has a well-defined large-$N$ expansion
such that its correlation functions factorize in the similar way as
those in the matrix model presented here, the disorder, as it is normalized
so far, is just too weak to cause any finite corrections 
to connected correlation functions in the leading order of $N$. 
This analysis demonstrates that the procedure of implementing the 
replica method in AdS/CFT as we explained in Sec.~\ref{sec:replica_ads}
were in fact working properly.

\section{The sub-leading contributions and a new holographic replica method?}
\label{sec:sub-leading}
The field theory analysis not only demonstrates that the holographic method
gives rise to the correct results, but also provides
us a mean to look for the sub-leading effects.

\begin{figure}[ht!]
\begin{center}
\includegraphics[width=0.80 \textwidth]{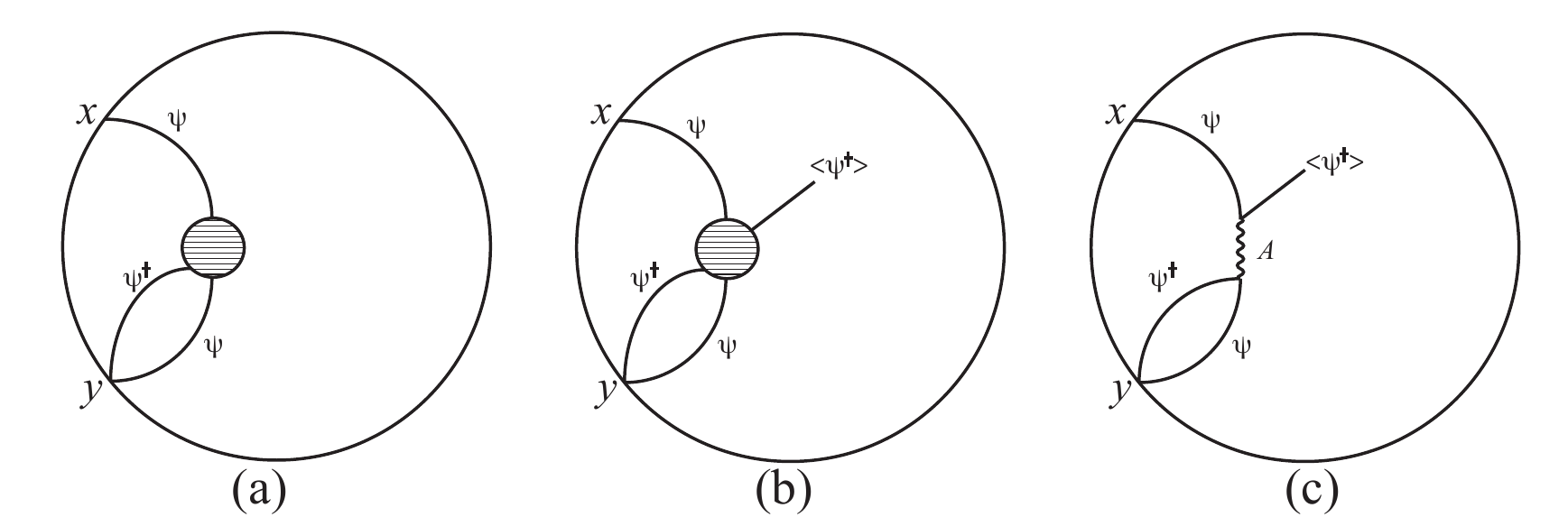}
\caption{\label{fig:deltavev1}Corrections to $\bar{\vev{O}}$ to the first order
of $f$. The vertex $y$ is integrated across the AdS boundary.}
\end{center}
\end{figure}

Let us study the order parameter, $\bar{\vev{O}}$, in the matrix model again.
We explained in the previous section that, at the order of $N^2$,
the corrections to it due to the double-trace deformation
vanish identically if we take the limit $n\rightarrow 0$.  
To the first order of $f$, the leading non-zero effects, described by the 
class of diagrams shown in Fig. \ref{fig:replica-leading}, can be written as
\[
\delta\vev{O(x)}
=\frac{f}{2}\int\ud^d y\, \lim_{y_{1,2}\rightarrow y}\vev{O(x)O(y_1)O(y_2)}^c\,.
\]
We had emphasized that splitting the double-trace
vertex and then taking the limit that they recombine was merely 
a formal step, which allowed us to easily visualize the difference between
the diagrams illustrated in Fig.~\ref{fig:witten-leading} and Fig.~\ref{fig:replica-leading}
and understand why the latter are relatively suppressed.
The explicit calculation in Appendix \ref{app:magic} presents an example where
the exact meaning of such a limit is given.
However, if we do take the above expression seriously, we may 
translate such correlation functions back to the AdS side 
using the standard dictionary. The result 
might be depicted by the Witten diagram shown in Fig. \ref{fig:deltavev1} (a).
The vertex on the boundary at $y$ should be understood as
the limit of two vertices $y_1, y_2\rightarrow y$, which explains
the two boundary-to-bulk correlators attached there, and is integrated
across the AdS boundary.  Formally, one may think of it as an insertion 
of a double trace operator on the CFT side. 
The circular blob in the bulk represents any possible diagrams 
that connect the two sides.

\begin{figure}[ht!]
\begin{center}
\includegraphics[width=0.50 \textwidth]{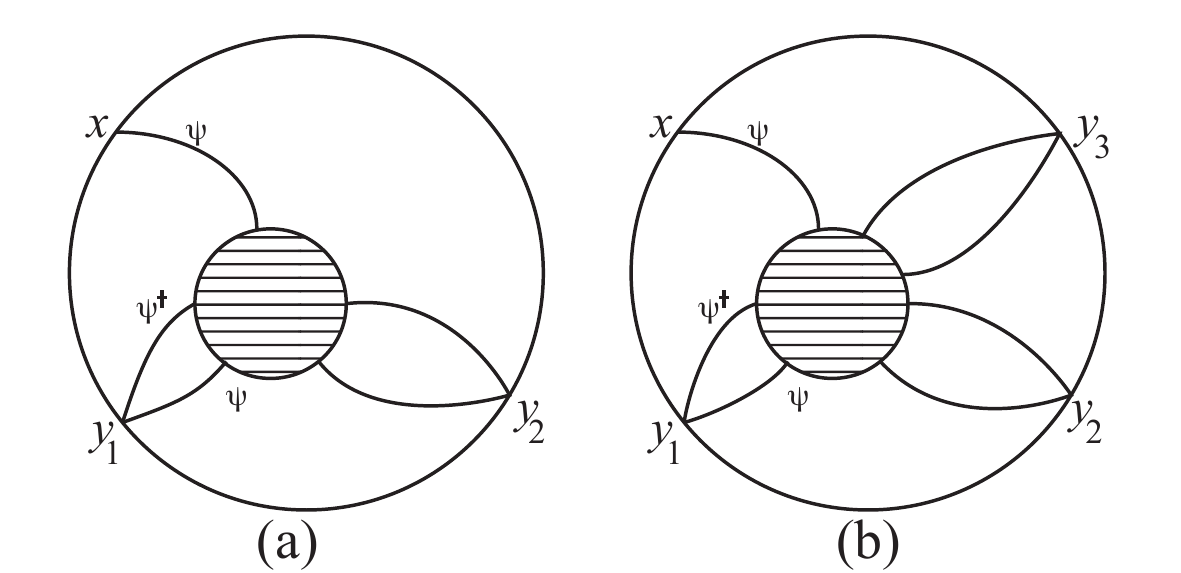}
\caption{\label{fig:deltavev2nd}Higher order corrections to $\bar{\vev{O}}$.
Vertices $y_{1,2,3}$ are integrated across the AdS boundary.}
\end{center}
\end{figure}

Similarly, one could also consider the higher order corrections
in terms of $f$. Again, let us take the
limit $y_i\rightarrow y$ literally and translate the
``super connected'' correlation functions to the AdS side,
we find that one only needs to insert more ``double-vertices'' on the AdS boundary
and integrate them all across it, as illustrated
in Fig. \ref{fig:deltavev2nd}. All such higher order corrections
are further suppressed by additional factors of $N^2$ though, so 
it is not quite useful to evaluate them if one does not
at the same time include all other competing contributions such as
the loop diagrams in the AdS bulk.

In passing, we mention an observation that follows immediately from
the above analysis. When the bulk Lagrangian is described by 
Eq.~\eqref{eq:ads_replica} and the disorder is coupled to
the operator $O$, to the leading order of $f$, the correction to 
the condensate $\bar{\vev{O}}$ only shows up if the system is in 
the superconducting phase, i.e. $\vev{O}\ne 0$, a result fully
expected of course. This is because the Witten diagram in Fig.~\ref{fig:deltavev1} (a)
can not be completed without breaking the $U(1)$ gauge symmetry.
Only if $\psi$ has a non-zero background profile, such diagram 
can be realized, as shown in Fig. \ref{fig:deltavev1} (b),
if one of the legs is connected to the background.
Fig.~\ref{fig:deltavev1} (c) gives a more explicit realization.
All those diagrams are accompanied by their conjugate of course.

\begin{figure}[ht!]
\begin{center}
\includegraphics[width=0.80 \textwidth]{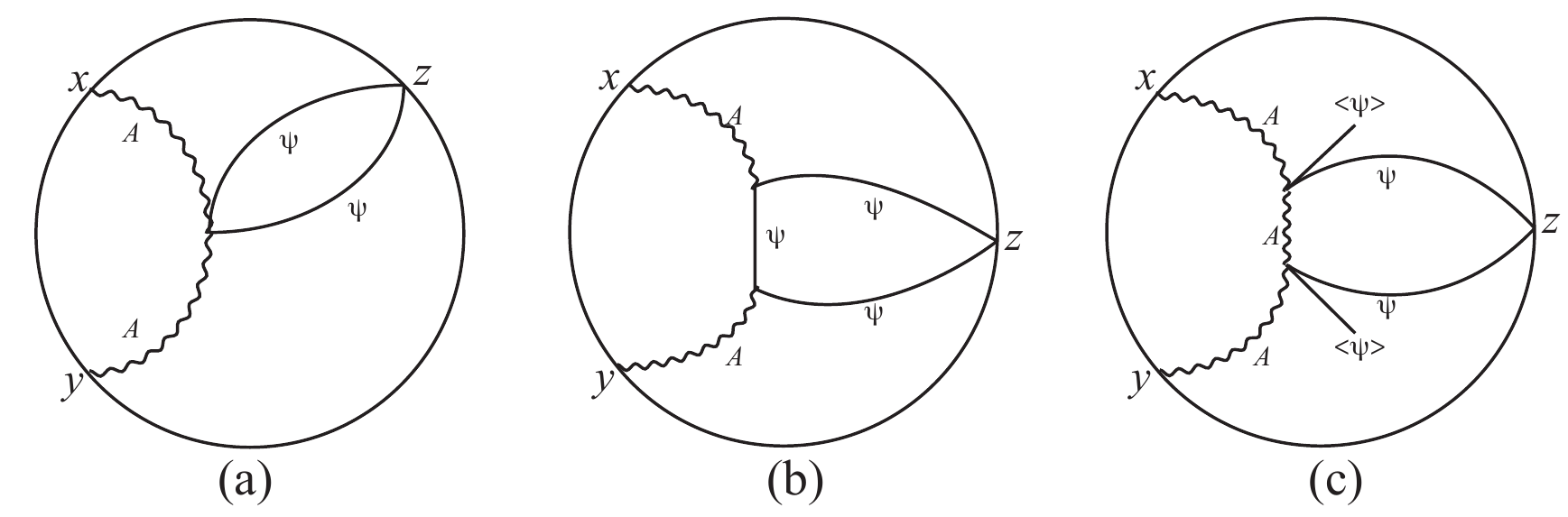}
\caption{\label{fig:delta2pt}The corrections to the averaged boundary-to-boundary
correlator for the gauge field to the first order of $f$.  The vertex $z$
is integrated across the AdS boundary.}
\end{center}
\end{figure}

We could also compute the corrections to the two-point correlation
functions perturbatively in $f$ in the same way. For example, 
the leading corrections to the gauge
field boundary-to-boundary propagator is given by the diagrams
shown in Fig. \ref{fig:delta2pt}.

The methods proposed here deserve some discussions. Treating
every insertion of a double-trace operator on the CFT side
by simply adding a ``double vertex'' on the AdS boundary for the
dual field seems superficial. If not in the context of the replica 
method, such a map is mistaken.
After all, whenever the number of replicas is greater than zero, 
the leading contributions are given by the diagrams show in 
Fig.~\ref{fig:witten-leading}, not at all included in the formalism 
suggested here. However, if one is only interested in the limit
$n\rightarrow 0$, the missing contributions all disappear, and
it becomes possible to described all the surviving contributions
by the simple Witten diagrams mentioned.

Some of the diagrams shown in Fig.~\ref{fig:deltavev1}
are similar to the 1-loop diagrams considered
in \cite{Hung:2010pe}, but milder since only one of the 
vertices of the ``loop'' needs to be integrated throughout the 
AdS bulk. They will not suffer from the extra ``volume'' divergences 
discussed in \cite{Hung:2010pe} for this reason. It is only the
$d$-dimensional Minkovski part of the loop integrals that may
need regularizations and the standard QFT methods in flat spacetime should suffice.

Apart from possible regularizations needed, the true weakness of the
procedure outlined here is that all higher order corrections are suppressed
by more powers of $N$ and it would be pointless to evaluate them
if one does not include all the other competing
effects at the same time. Let us mention, however, a potentially much more
interesting scenario.  It is conceivable to make the 
effects of the disorder stronger so they show up at the leading order of
the large-$N$ expansion, i.e. $O(N^2)$.  To do so, we just need to make the disorder
stronger as $N\rightarrow \infty$, leading to a double-trace deformation
with a different normalization:
\begin{equation}\label{eq:new_normal}
\delta\mathcal L=\frac{N^2 f}{2} O^2\,.
\end{equation}
Formally, we can achieve this goal in the matrix model
if we just let the coupling between
$V$ and $O$ depend on the gauge coupling $g$.  More specifically,
we may put the same overall factor of $g^{-2}$ in front of the coupling of 
$V$ and $O$ so it reads
\[
\delta L_V=\frac{1}{g^2} V(x) O(x)\,.
\]
Integrating out $V$ leads to a deformation 
term $\delta \mathcal L=\frac{f}{2 g^4}O^2$
and in the t' Hooft limit it becomes $\delta \mathcal L=\frac{N^2f}{2\lambda^2} O^2$.
After redefining $\lambda^{-2} f\rightarrow f$, we find \eqref{eq:new_normal}.
Equivalent, we may change the Gaussian distribution profile for
$V$ so that the width becomes $f N^2$.
In any case, it amounts to choose a different normalization for
the single-trace operator coupled to the disorder.
By giving $f$ an extra factor $N^2$, the leading diagrams shown in 
Fig.~\ref{fig:replica-leading} are made $O(N^2)$.

Normally, a double-trace term normalized as in Eq.~\eqref{eq:new_normal}
is disastrous, because it destroys the large-$N$ expansion.
As we bring up the sub-dominant contributions to the 
leading order, we necessarily make all the leading ones, shown in 
Fig.~\ref{fig:witten-leading}, too large simultaneously. 
As $N\rightarrow\infty$, they totally over-dominate any other physics. 
If one consider the higher order corrections in terms of $f$, situation 
only becomes worse and worse, essentially making the large-$N$ limit singular. 
Consequently, one does not expect to find a holographic dual 
describable by a semi-classical field theory.

In the current context, however,
we are rescued by the magic power of the replica method.
Because we are only interested in the limit $n\rightarrow 0$ 
and those bad-behaved diagrams are known to vanish in this limit, it 
is hopeful that the said disease can be cured.
Indeed, the diagrams as shown in Fig.~\ref{fig:witten-leading} are at best of the 
order $n N^4$ as one can easily verify in the example given in
Appendix~\ref{app:magic}.  Therefore, it suffices if we somehow restrict
ourselves in the corner of the parameter space, such that, as
$N\rightarrow \infty$, $\tilde n=n N^2$ is kept finite.
That way, the diagrams of the order $n N^4$ become $O(N^2)$
as well. It is not hard to check the pattern is maintained
when higher order corrections in terms of $f$ are considered.
Setting $n=0$ leaves with us only the diagrams
similar to those illustrated in Fig.~\ref{fig:replica-leading}, but
leaving $\tilde n\sim O(1)$ allows both diagrams in Fig.~\ref{fig:witten-leading} 
and Fig.~\ref{fig:replica-leading} to contribute at the same order.

Interestingly, with the normalization in Eq.~\eqref{eq:new_normal},
it also happens automatically that the higher order contributions in terms of the
deformation strength $f$ are no longer relatively suppressed. All of them
become $O(N^2)$ and contribute at the leading order in the large-$N$
expansion simultaneously, allowing themselves to stand out and be separated 
from other competing effects, such as those of the non-planar diagrams,
or correspondingly loops in the AdS bulk.  
In this theory, both $f$ and $\tilde n$ can be considered as an
additional 't~Hooft coupling.

This observation also implies that it becomes both necessary and meaningful to 
sum all the contributions of the double-trace deformation
to all orders of $f$, which is out of our reach at
the moment unfortunately.  It is plausible 
that a modified AdS/CFT correspondence, yet unknown, may exist that 
capture all the contributions non-perturbatively.  Somehow the
dual theory must incorporate $\tilde n$ in a non-trivial way
because only when it is kept $O(1)$, a consistent large-$N$ 
expansion can be defined.

\section{Discussion and Conclusions}
\label{sec:conclusion}
While we used the matrix theory as a toy model to probe 
the effect of multi-trace deformations perturbatively on the field
theory side, the discussion in the previous sections solely relies
on the essential property that the correlation functions 
factorize in the large-$N$ expansion. We believe that the found sub-leading
contributions remain the same for general theories with holographic duals.

For realistic physical questions, however,
one must restrict himself to the time-independent random disorders, 
which is equivalent to giving the disorder
strength $f$ a factor of $\delta(\omega)$ where $\omega$ is the frequency of the disorder.
For numerical calculations, one must then smear out this delta-function somewhat,
replace it by a sharp Gaussian-like profile, and take the limit properly in the end.

In our discussions above, we had not specified the spacetime signature.
What we found obviously applies in the Euclidean space where
the correlation functions are defined as written. In the real-time
formalism, however, various type of correlation functions can be defined,
and physically the retarded ones are the most interesting.
Our discussions above should apply to all correlation functions, since the
counting of the powers of $N$ is a separate issue. On the field theory side,
different correlation functions correspond to different choice of the
time-contour in the path-integral, and, on the AdS side, the
choice amounts to specifying different boundary conditions for the
mode solutions at the horizon \cite{Son:2002sd, Herzog:2002pc, Skenderis:2008dg, Iqbal:2009fd}, 
and the discussions presented above remain unchanged.

It is a bit disappointing that the method proposed here for
evaluating the non-vanishing sub-leading effect
is only perturbative in terms of the disorder strength $f$.  In fact, it is simply
pointless to evaluate the higher order corrections on their own, since being more and
more suppressed in the large-$N$ expansion, they are contaminated by
more and more competing contributions not discussed in this paper.

A much more interesting scenario is found if we decide
to increase the disorder strength such that, in the replica method, it
leads to a double-trace deformation normalized as in Eq.~\eqref{eq:new_normal}. 
In an ordinary CFT, such a normalization is pathological,
since connected diagrams in the deformed theory may carry arbitrarily 
higher powers of $N$, totally invalidating the large-$N$ expansion.  However, 
as we put the formalism in the context of the replica method, the situation becomes
very different, because there emerges a new parameter that is small, namely,
the number of the replicas. We found that the problematic diagrams that 
may carry higher powers of $N$ must always be proportional
to $(n N^2)^\# N^2$. So, focusing on the limit $n\rightarrow 0$ allows
us to consider this theory more seriously.
In fact, as long as we can keep $\tilde n=n N^2$ finite as 
$N\rightarrow \infty$, all the leading connected diagrams are
$O(N^2)$ and a consistent large-$N$ expansion can be defined,
in which $\tilde n$ may be thought of as a new 't~Hooft coupling.  
In the meanwhile, higher order contributions in terms of $f$ are also
no longer suppressed, and they all contribute at the order of $N^2$. 
Therefore, at every order of $f$, the leading contributions due to 
the multi-trace operator can be cleanly separated from 
other corrections such as those given by the non-planar diagrams.
It would be extremely interesting, if one could 
find a new holographic model for such a theory
that captures all the effects non-perturbatively.
This is, of course, a novel regime since the duality may only work properly
if we can keep the number of the fields, $n$, to be $O(N^{-2})$ as 
$N\rightarrow\infty$, something that remains unclear to us how to
achieve but is surely worth further investigations.

\section*{Acknowledgement}
We thank Ling-Yan Hung for collaboration in the early part of the 
work, and Cliff Burgess, Sung-Sik Lee, and Allan Bayntun for the very helpful
discussions and suggestions.
Research at Perimeter Institute is supported by the 
Government of Canada through Industry Canada and by the 
Province of Ontario through the Ministry of Research \& Innovation.

\appendix
\section{The replica method for free theories}\label{app:free}
The simplest example that the replica method can be explicitly implemented
is a free theory where the operator coupled to the disorder is the
fundamental field itself. It is unnecessary to elaborate
such a simple example if not because we find that the holographic calculation
surprisingly resembles the free theory in various ways.

Let us consider the action $S_0=-\frac{1}{2}K X^2$, where we are being
schematic by suppressing space-time integrals and denote the kinetic operator
simply by $K$.  Let it be coupled to a random potential $V$ so we have
$S=S_0+V X$ and let $V$ has a distribution given by $P[V]=\exp\{-\frac{f}{2}V^2\}$.
By the replica method, we should consider the following action:
\[
S=-\frac{1}{2}\sum_i^n K X_i^2+\frac{f}{2}\left(\sum_i^n X_i\right)^2\,.
\]
To deal with the mixing term, we define
$\tilde X_i=\sum_j^n a_{ij} X_j$, where $a_{ij}$ is the matrix defined
in Sec.~\ref{sec:replica}. In terms of $\tilde X_i$ the
action is diagonalized and we have
\[
S=-\frac{1}{2}\sum_i^n K\tilde X_i^2+\frac{nf}{2}\tilde X_n^2\,.
\]
So we easily find
\[
\bar{\vev{X}}=\lim_{n\rightarrow 0}\frac{1}{n}\vev{\sum_i^n X_i}
=\lim_{n\rightarrow 0}\vev{\tilde X_n}=0\,.
\]
Slightly more interestingly, we have
\[
\vev{\tilde X_n \tilde X_n}=\frac{1}{K-nf}\,,\quad
\vev{\tilde X_{n-1}\tilde X_{n-1}}=\frac{1}{K}\,.
\]
By the formulae Eq. \eqref{eq:2pt_tilde}, we find
\[
\begin{split}
\bar{\vev{XX}}=&\lim_{n\rightarrow}\frac{1}{n}
\left[\vev{\tilde X_n\tilde X_n}+(n-1)\vev{\tilde X_{n-1}\tilde X_{n-1}}\right]
=\frac{1}{K}+\frac{f}{K^2}\,,\\
\bar{\vev{X}\vev{X}}=&\left[\vev{\tilde X_n\tilde X_n}-
\vev{\tilde X_{n-1}\tilde X_{n-1}}\right]
=\frac{f}{K^2}\,,
\end{split}
\]
but the connected correlation function, given by the
difference of the two expressions above, is independent of $f$ and
\[
\bar{\vev{XX}^c}=\frac{1}{K}\,.
\]
This is not at all surprising. Since the action is quadratic,
the connected two-point function for $X$ is independent to the linear
term $XV$, so for any chosen $V$, the correlator remains to be $1/K$, 
and no wonder the averaged value is the same. The essential point
that makes such simple calculations possible is that the system can
be easily diagonalized by the rotation $a_{ij}$.
We see that the same holds true for correlation
functions in the context of the holographic replica method.

\section{An explicit calculation that verifies Eq.~\ref{eq:C_formula}}
\label{app:magic}
While Eq. \eqref{eq:C_formula} is general, it is not fully self-evident.
We present an explicit calculation in the case of a 
single double-trace insertion, which not only verifies the Eq.~\eqref{eq:C_formula}
but also helps to clarify the actual meaning of the final expression given there.

Let us consider, for example, the leading order correction to $\vev{O}$
given by the integral:
\[
\delta\vev{O_1(x)}^\circ=\frac{f}{2}\lim_{n\rightarrow 0}\frac{1}{n}\int\ud^d y
\vev{O_1(x)\sum_{i=1}^n O_i(y) \sum_{j=1}^n O_j(y)}^\circ
\]
Here, all the correlation functions on the right-hand side are evaluated in the limit 
$f=0$, and so all the replicas decouple. Clearly, we have 
\[
\vev{O_i(a) O_j(b)}^\circ=Z^n\vev{O_i(a)}\vev{O_j(b)}
\]
for $\forall i\ne j$. Here $Z=\int\uD X \exp\{S[x]\}$ is the partition function for
the single-copied theory, and VEVs on the right-hand side of the above 
equation are normalized.  So to the leading order of $f$, we find
\[
\begin{split}
\delta\vev{O_1(x)}^\circ=\frac{f Z^n}{2}\lim_{n\rightarrow 0}\frac{1}{n}\int\ud^d y
&\left[
\vev{O_1(x)O_1(y) O_1(y)}+(n-1)\vev{O_1(x)}\vev{O_2(y) O_2(y)}\right.\\
&+2(n-1)\vev{O_1(x)O_1(y)}\vev{O_2(y)}\\
&\left.+(n-1)(n-2)\vev{O_1(x)}\vev{O_2(y)}\vev{O_3(y)}
\right]\\
&=\frac{f}{2}\int\ud^d y\vev{O(x)O(y)O(y)}^c,
\end{split}
\]
where
\[
\vev{ABC}^c\equiv\vev{ABC}-\vev{AB}\vev{C}-\vev{A}\vev{BC}-\vev{AC}\vev{B}
+2\vev{A}\vev{B}\vev{C}
\]
is precisely the totally connected correlation functions
if $A$, $B$ and $C$ are all separated.

\end{document}